\documentclass[aps,pra,preprint,showpacs]{revtex4}

\usepackage{graphicx}
\usepackage{amsmath}
\usepackage{bm}

\begin{document}

\title{Ps-atom scattering at low energies}

\author{I. I. Fabrikant}
\affiliation{Department of Physics and Astronomy, University of Nebraska,
Lincoln, Nebraska 68588-0299, USA}
\author{G. F. Gribakin}
\affiliation{School of Mathematics and Physics, Queen's University Belfast,
Belfast BT7 1NN, Northern Ireland, United Kingdom}

\date{\today}

\begin{abstract}
A pseudopotential for positronium-atom interaction, based on
electron-atom and positron-atom phase shifts, is constructed, and the
phase shifts for Ps-Kr and Ps-Ar scattering are calculated. This approach
allows us to extend the Ps-atom cross sections, obtained previously in the
impulse approximation [Phys. Rev. Lett. {\bf 112}, 243201 (2014)],
to energies below the Ps ionization threshold.
Although experimental data are not available in this low-energy region,
our results describe well the tendency of the measured cross sections to drop 
with decreasing velocity at $v<1$~a.u.
Our results show that the effect of the Ps-atom van der Waals interaction is 
weak compared to the polarization interaction in electron-atom
and positron-atom scattering. As a result, the Ps scattering length for
both Ar and Kr is positive, and the Ramsauer-Townsend minimum is not observed 
for Ps scattering from these targets. This makes Ps scattering quite different 
from electron
scattering in the low-energy region, in contrast to the intermediate energy 
range from the Ps ionization threshold up to $v\sim 2$~a.u., where the
two are similar.
\end{abstract}

\pacs{34.80.-i, 36.10.Dr}

\maketitle

\section{Introduction}

Recently observed similarities between the positronium scattering and
the electron
scattering from a number of atoms and molecules \cite{Bra10,Bra10a,Bra12}
in the intermediate energy range suggest
that both processes are largely controlled by the same interactions. When
plotted as a function of the projectile velocity, the electron and Ps cross
sections are very close and even show similar resonancelike features.
This similarity was explained recently \cite{Fab14}
using the impulse approximation.
It was shown that above the Ps ionization threshold, the Ps-$A$ interaction is 
mainly controlled by the $e^-$-$A$ scattering amplitude, and the $e^-$-$A$ 
exchange contributes mostly to this amplitude in the intermediate energy range.
It is clear, however, that at higher energies, the role of the exchange
interaction becomes less significant, and the similarity between Ps-$A$ and
$e^-$-$A$ scattering should gradually disappear.
On the other hand, at
lower energies, long-range interactions between the projectile and the target 
play a significant role, and they are substantially different for $e^-$-$A$
and Ps-$A$ interactions. In the former case it is the polarization potential 
decreasing as $r^{-4}$ at large distances $r$, and in the latter case the van 
der Waals interaction decreasing as $r^{-6}$.

The impulse approximation for Ps-$A$ scattering \cite{Fab14} produces
very large cross sections below the ionization threshold because of the
dominance of the $e^+$-$A$ scattering amplitude. This growth of the Ps cross 
section is unphysical, since the large  $e^+$-$A$ amplitude is due to
the effects of positron-atom polarization and virtual Ps formation,
both of which are absent in Ps-$A$ scattering.
Since the impulse approximation breaks down at energies below the ionization 
threshold,
alternative methods and approximations, such as close-coupling and 
static-exchange, should be used in this energy range.
Blackwood {\it et al.} \cite{Bla02} performed close-coupling calculations of 
Ps scattering from noble-gas atoms.
These calculations allowed for the distortion and break-up of Ps, but kept
the target ``frozen'', i.e., they neglected any excitations of the target. 
Virtual target excitations are known to be very important 
in low-energy electron- and positron-atom collisions, where they
can be described in terms of the polarization interaction.
For low-energy Ps-atom collisions, they give rise to the van der
Waals interaction, which can be incorporated
by extending the close-coupling calculations to include the virtual
excitations of the target. Such calculations have been performed for Ps 
collisions with the hydrogen atom \cite{Cam98,Bla02a} and would be an 
ultimate goal in the problem of Ps-atom collisions. However,
they are very challenging computationally for complex atoms.

In this paper we develop an alternative low-energy method based on the use 
of the electron and positron scattering phase shifts, similar to the impulse 
approximation. It involves constructing model potentials that reproduce these 
phase shifts, and then adding them to describe the Ps-atom interaction. While 
this procedure is straightforward for positron scattering, the situation with 
electrons is more
complicated. Due to the Pauli exclusion principle, the effective potential
for the electron depends on its orbital angular momentum $l$, i.e., it becomes 
a pseudopotential \cite{Pas83}. When such a pseudopotential is averaged over 
the electron density distribution in the Ps atom, it becomes a nonlocal 
operator.

Another difficulty is related to inclusion of the long-range interaction.
Accurate low-energy electron and positron scattering phase shifts contain
contributions of the atomic polarization potential. This potential is
attractive for both electrons and positrons and behaves as $-\alpha /2r^4$
at large distances, where $\alpha $ is the atomic dipole polarizability.
(We use atomic units throughout.) An effective Ps-$A$ potential including
such contributions would behave as $-\alpha /r^4$, which is physically
incorrect, as the dominant long-range Ps-$A$ interaction is the van der Waals
interaction $-C_6/R^6$. The latter potential results from the many-body
Ps-atom, rather than single-particle (i.e., electron-atom or positron-atom)
dynamics. It can be obtained by including the \textit{two-body} polarization
potential (see, e.g., \cite{Mit01}) in the Hamiltonian, which gives the total
polarization interaction at large distances as
\begin{equation}\label{tb_pol}
V_{\rm pol}({\bf r}_e,{\bf r}_p)=-\frac{\alpha }{2r_e^4}-\frac{\alpha }{2r_p^4}
+\frac{\alpha \,{\bf r}_e\cdot {\bf r}_p }{r_e^3r_p^3}
\end{equation}
where ${\bf r}_e$ and ${\bf r}_p$ are the electron and positron position
vectors, respectively, relative to the target. Averaging of this potential over
the electron 
and positron density distribution in Ps does lead to an effective van der Waals
interaction \cite{comment}.

Alternatively, one can construct the positron-atom and electron-atom 
pseudopotentials using static (static-exchange) phase shifts for the
positron (electron), i.e., neglecting the polarization. The Ps-atom van der 
Waals interaction can then be added, e.g., in the form
\begin{equation}\label{vdW}
V_W(R)=-\frac{C_6}{R^6}\left\{1-\exp[-(R/R_c)^8]\right\} ,
\end{equation}
where $C_6$ is the van der Waals constant and $R_c$ is a cutoff radius.
The $C_6$ values for Ps-atom pairs are known, e.g., from the London formula 
\cite{London}, which gives $C_6=152$~a.u. for Kr and $C_6=104.5$~a.u. for 
Ar. (These values are close to the estimates obtained in Ref. 
\cite{MB03}.) In contrast, the cutoff parameter $R_c$ cannot be
determined rigorously, but the phase shifts and cross
sections are sensitive to its choice. A similar problem is encountered
when using the polarization interaction (\ref{tb_pol}), which also requires 
a cut-off at small distances.
In the present calculations the radius $R_c$ is set by 
requiring that the cross sections given by the pseudopotential
method merge smoothly with the elastic cross section calculated in the
impulse approximation above the Ps ionization threshold \cite{Fab14}.

The rest of the article is organized as follows. First we discuss the
construction of the pseudopotentials for $e^+$ and $e^-$
scattering from the static (static-exchange) phase shifts, and the derivation 
of the pseudopotential for Ps-atom scattering. We then present the results for 
Ps-Kr and
Ps-Ar scattering and discuss the low-energy behavior of the cross sections.
For both atoms, the scattering length is positive, which implies effective
repulsion at low energies and rules out the existence of the Ramsauer-Townsend 
minimum.

\section{Theory}
\subsection{Pseudopotentials}\label{subsec:pp}

We choose the positron-atom pseudopotential in the form
\begin{equation}\label{V_p}
V_p(r)=\frac{Z_p}{r}e^{-\alpha _pr},
\end{equation}
which represents the static $e^+$-$A$ repulsion, and where
$Z_p$ and $\alpha _p$ are fitting parameters. They are obtained by fitting
the $s$-, $p$- and $d$-wave scattering phase shifts in the potential
(\ref{V_p}) to the positron scattering phase shifts in the static field
of the ground-state atom calculated in the Hartee-Fock approximation
(see Sec.~\ref{subsec:fit}). The parameter $Z_p$ plays the role of an
effective nuclear charge. It can be different from the actual nuclear
charge, since a low-energy positron does not penetrate deep
into the atom.

The effective static-exchange potential for the electron is chosen as
\begin{equation}\label{V_el}
V_e(r)=-\frac{Z_e}{r}e^{-\alpha _er}+\frac{B}{r^n}e^{-\beta r},
\end{equation}
where the second term represents repulsion due to the Pauli
exclusion principle. This effect depends on the orbitals occupied in the atomic
ground state, hence it is $l$-dependent. We also found that in general the
{\it ab initio} static-exchange (Hartree-Fock) phase shifts cannot be
reproduced using $Z_e=Z_p$. Therefore, we regard all parameters in
Eq.~(\ref{V_el}) as $l$-dependent. Formally this means that the effective
electron-atom potential is a nonlocal operator with the kernel
\begin{equation}
V_e({\bf r},{\bf r}')=\frac{1}{r^2}\delta(r-r')\sum_{lm}V_l(r)Y_{lm}^*(\hat{\bf r})
Y_{lm}(\hat{\bf r}'),
\end{equation}
where $V_l(r)$ are potentials given by Eq. (\ref{V_el}). It is convenient
to rewrite this expression as
\begin{equation}
V_e({\bf r},{\bf r}')=-V_p(r)\delta({\bf r}-{\bf r}')+
\frac{1}{r^2}\delta(r-r')\sum_{lm}v_l(r)Y_{lm}^*(\hat{\bf r})
Y_{lm}(\hat{\bf r}'),
\end{equation}
where
\begin{equation}
v_l(r)=V_l(r)+V_p(r).
\label{vl_1}
\end{equation}
Since the ``direct'' part of the potential $V_l(r)$ is close in magnitude,
but opposite in sign to $V_p(r)$, $v_l(r)$ represents mainly the exchange
interaction between the electron and the atom.

The Ps-atom pseudopotential can be now written as
\begin{equation}
V_{\rm Ps}({\bf r}_e,{\bf r}'_e,{\bf r}_p,{\bf r}'_p)=
V_p(r_p)\delta ({\bf r}_p-{\bf r}'_p)-V_p(r_e)
\delta ({\bf r}_e-{\bf r}'_e)+
\frac{1}{r_e^2}\delta(r_e-r'_e)\sum_{lm}v_l(r_e)Y_{lm}^*(\hat{\bf r}_e)
Y_{lm}(\hat{\bf r}'_e).
\label{V_Ps}
\end{equation}
In the static approximation we average this potential over the electron and
positron density distribution in Ps given by $|\Phi(\bm{\rho})
|^2$, where $\Phi(\bm{\rho})$ is the Ps ground-state wave function and
$\bm{\rho}$ is the relative $e^-$-$e^+$ coordinate. The relations
between ${\bf r}_e$, ${\bf r}_p$ and $\bm{\rho}$ are
\[
{\bf r}_e={\bf R}+\bm{\rho}/2,\quad {\bf r}_p={\bf R}-\bm{\rho}/2,
\]
where {\bf R} is the position of the Ps center of mass relative to the target.

The average of the local part of the pseudopotential (\ref{V_Ps}) reduces to
the integral
\begin{equation}
\int \left[V_p({\bf R}-\bm{\rho}/2)-V_p({\bf R}+
\bm{\rho}/2)\right]|\Phi(\bm{\rho})|^2
d\bm{\rho},
\end{equation}
which vanishes because the integrand is parity odd. This corresponds to
a well-known fact that the static potential for the Ps-$A$ interaction is zero.
The remaining nonlocal part in Eq. (\ref{V_Ps}) contains a strong repulsive
core, and to make the calculations more tractable, it is convenient to
represent $v_l(r_e)$ as
\[
v_l(r_e)=v_{\rm loc}(r_e)+u_l(r_e),
\]
where the $l$-independent part $v_{\rm loc}(r_e)$ contains the major repulsive
contribution, and $u_l(r_e)$ accounts for the remaining $l$-dependent part.
The averaging procedure is then reduced to averaging of the Ps pseudopotential
\begin{equation}\label{Vtilde_Ps}
\tilde{V}_{\rm Ps}({\bf r}_e,{\bf r}'_e)=
v_{\rm loc}(r_e)\delta({\bf r}_e-{\bf r}'_e)+
\frac{1}{r_e^2}\delta(r_e-r'_e)\sum_{lm}u_l(r_e)Y_{lm}^*(\hat{\bf r}_e)
Y_{lm}(\hat{\bf r}_e').
\end{equation}
The average of the local part of this potential gives the local Ps-atom
potential
\begin{equation}\label{Vav_loc}
v_{\rm av}(R)=\int v_{\rm loc}({\bf R}+\bm{\rho}/2)
|\Phi(\bm{\rho})|^2d\bm{\rho},
\end{equation}
while averaging the nonlocal part [second term in Eq.~(\ref{Vtilde_Ps})] gives
a nonlocal contribution to the Ps-$A$ interaction,
\begin{equation}\label{Vav_nonloc}
V({\bf R},{\bf R}')=\sum_{lm}\int \frac{1}{r^2}\delta(r-r')
u_l(r)Y_{lm}^*(\hat{\bf r})Y_{lm}(\hat{\bf r}')
|\Phi(\bm{\rho})|^2d\bm{\rho}
\end{equation}
where ${\bf r}={\bf R}+\bm{\rho}/2$ and ${\bf r}'={\bf R}'+\bm{\rho}/2$.

In performing these integrations, it is more convenient to switch to
the integration variable ${\bf r}$. Since the result depends only on the
absolute magnitude of the vectors ${\bf R}$ and ${\bf R}'$, and on the angle
between them, the integration  in Eq.~(\ref{Vav_nonloc}) can be performed
in the coordinate system with the polar axis along the vector
\[
{\bf s}={\bf R}'-{\bf R}.
\]
We then have
\begin{equation}
V({\bf R},{\bf R}')=8\sum_{l}\frac{2l+1}{4\pi}\int \frac{1}{r^2}\delta
(r-r')u_l(r)P_l(\cos\theta_{{\bf rr}'})|\Phi(2({\bf r}-{\bf R}))|^2
d{\bf r},
\label{int1}
\end{equation}
where ${\bf r}'={\bf r}+{\bf s}$ and $\theta_{{\bf rr}'}$ is the angle
between {\bf r} and ${\bf r}'$.

Integration over the polar angle $\theta$ in this coordinate system eliminates
the $\delta $-function (which ensures $r=r'$) and fixes the angles:
\[
\cos\theta=-\frac{s}{2r}, \quad \cos\theta_{{\bf rr}'}=
\frac{r+s\cos\theta}{|{\bf r}+{\bf s}|}=1-\frac{s^2}{2r^2}.
\]
It also introduces a factor $1/s$ since
\[
\frac{d(r-|{\bf r}+{\bf s}|)}{d\cos\theta }=-s
\]
for $\cos\theta=-s/2r$.

The Ps ground-state density is expanded in spherical harmonics as
\begin{equation}\label{eq:Phi2}
|\Phi(2|{\bf r}-{\bf R}|)|^2=\frac{1}{8\pi}e^{-2|{\bf r}-{\bf R}|}=
\frac{1}{8\pi}\sum_{l'=0}^\infty F_{l'}(r,R)(2l'+1)P_{l'}(\cos\theta_{\bf r R}),
\end{equation}
where the expression for $F_l(r,R)$ is given in Appendix~\ref{app:F}
(see also Appendix B in Ref. \cite{Dzu96}).

The Legendre polynomial $P_{l'}(\cos\theta_{\bf r R})$ is the only
part of expansion (\ref{eq:Phi2}) which depends on the azimuthal
angle $\phi$ in the integrand of Eq.~(\ref{int1}) ($\theta_{{\bf
rr}'}$ does not depend on $\phi$). Therefore we can perform
integration over $\phi$ as
\[
\int _0^{2\pi }P_{l'}(\cos\theta_{\bf r R})d\phi=2\pi P_{l'}
(\cos\theta)P_{l'}(\cos\theta_R),
\]
where
\[
\cos\theta_R=\frac{R'\cos\Theta-R}{s}
\]
and $\Theta$ is the angle between ${\bf R}$ and ${\bf R}'$.

Thus, we obtain the nonlocal part of the Ps-atom potential as
\begin{equation}
V({\bf R},{\bf R}')=\frac{1}{2\pi s}\sum_{l\,l'}(2l+1)(2l'+1)
P_{l'}(\cos\theta_R)\int _0^\infty P_{l}\left(1-\frac{s^2}{2r^2}\right)
P_{l'}\left(-\frac{s}
{2r}\right)F_{l'}(r,R)u_{l}(r)dr.
\label{vnonlocal}
\end{equation}
It is convenient to expand this expression in Legendre polynomials,
\[
V({\bf R},{\bf R}')=\frac{1}{RR'}\sum_{L=0}^\infty \frac{2L+1}{4\pi}
V_L(R,R')P_L(\cos\Theta),
\]
where
\[
V_L(R,R')=2\pi RR'\int _0^\pi V({\bf R},{\bf R}')
P_L(\cos\Theta)\sin \Theta
d\Theta. \]
Similarly, for the local part of the interaction potential, given by
Eq.~(\ref{Vav_loc}), we obtain
\[
v_{\rm av}(R)=4\int _0^\infty v_{\rm loc}(r)F_0(r,R)r^2dr.
\]
Substitution of the local and nonlocal potentials in the Schr\"odinger equation for 
the Ps-$A$ system yields a set of radial equations:
\begin{equation}\label{eq:Sch}
\frac{1}{2m}\frac{d^2f_L}{dR^2}+\left[E-v_{av}(R)-\frac{L(L+1)}
{2mR^2}\right]f_L(R)-\int V_L(R,R')f_L(R')dR'=0,
\end{equation}
where $ m=2$ a.u. is the Ps mass, and $f_L(R)$ is the radial part of the Ps
center-of-mass wave function for the orbital angular momentum $L$.

The sums in Eq. (\ref{vnonlocal}) converge fast. With a proper choice
of $v_{\rm loc}(r_e)$, the sum over $l$ can be truncated at
$l_{\max}=2$. Increasing $l_{\max}$ to 4 has almost no effect on the phase 
shifts in the velocity range up to 2 a.u. The sum over $l'$ converges if
$l'_{\max}\ge 6$.

\subsection{Fitting parameters}\label{subsec:fit}

The values of the parameters of the pseudopotentials in Eqs. (\ref{V_p}) and 
(\ref{V_el}), were determined by fitting the positron-atom and
electron-atom scattering phase shifts obtained in the static potential of
the atom calculated in the Hartree-Fock approximation \cite{ATOM}.

\begin{figure}[ht!]
\includegraphics[width=10cm]{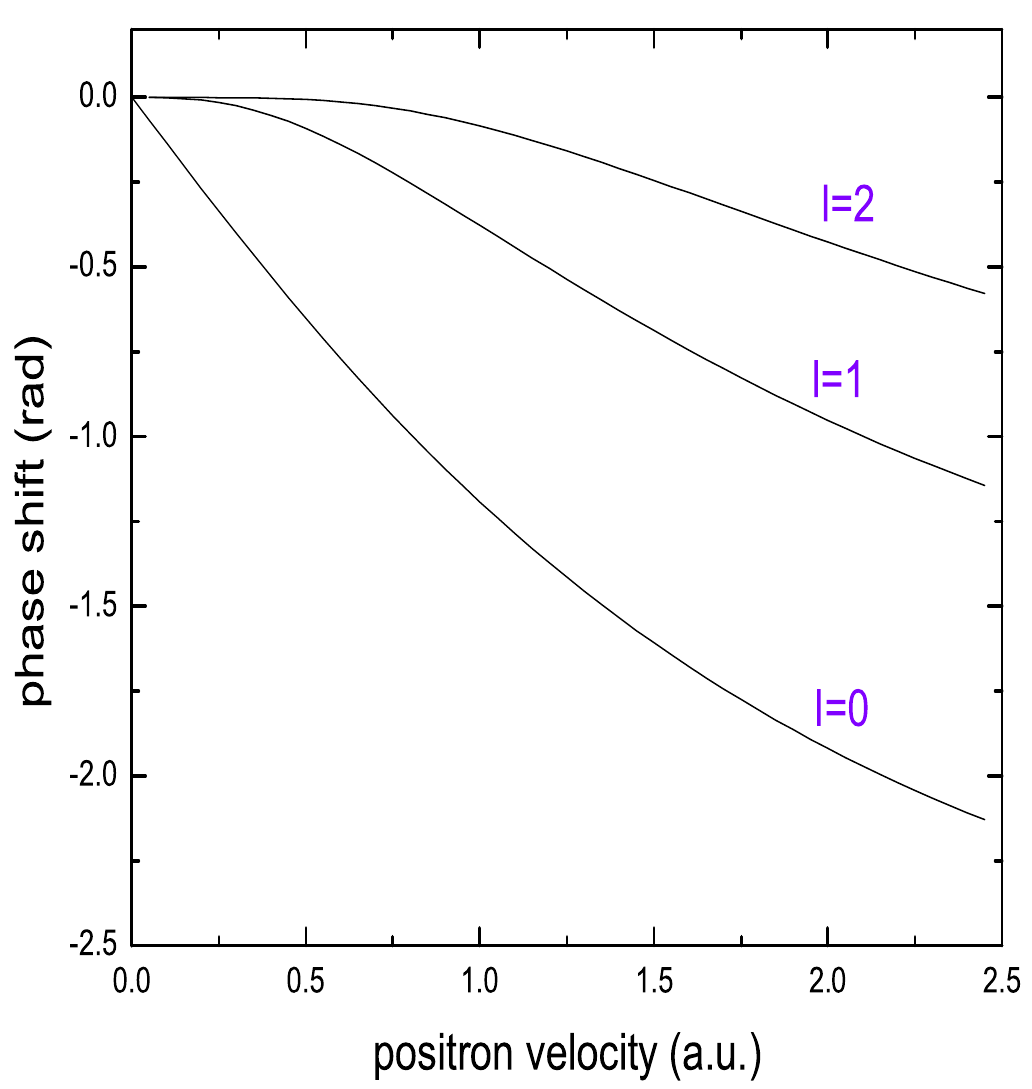}
%\picturelandscape{0}{pos_Ar.eps}
\caption{Positron scattering phase shifts for Ar obtained using the
static potential of the Hartree-Fock atom.}
\label{fig:posAr}
\end{figure}

As mentioned in Sec.~\ref{subsec:pp}, for the positron, the same pseudopotential 
can be used for  all partial waves.
The static-field scattering phase shifts for the positron on Ar
are shown in Fig.~\ref{fig:posAr}, and the corresponding sets of
parameters for Ar and Kr are given in Table~\ref{tb:parameters}. For positron 
velocities up to 2~a.u., the pseudopotential phase
shifts are within 1\% of the actual static-field phase shifts and
are indistinguishable from them on the scale of the plot.

\begin{table}[ht!]
\begin{ruledtabular}
\caption{Parameters of the positron and electron pseudopotentials, given by
Eqs.~(\ref{V_p}) and (\ref{V_el}), for Ar and Kr.}
\label{tb:parameters}
\begin{tabular}{ccccccc}
System & $l$ & $Z$ & $\alpha _{p,e}$ & $B$ & $n$  & $\beta$ \\
\hline
$e^+$-Ar & 0--4 & 18.06 & 1.95   &  --   &   --    &   --\\
\hline
$e^-$-Ar & 0   & 4.297 & 0.618  & 14.72 & 3  & 0.248 \\
         & 1   & 10.0  & 2.368  & 15.24 & 3  & 0.504 \\
         & 2   & 9.780 & 1.230  &  0    & -- &  --     \\
         & 3   & 12.48 & 1.544  &  0    & -- &  --     \\
         & 4   & 15.13 & 1.714  &  0    & -- &  --     \\
\hline
$e^+$-Kr & 0--4 & 20.79 & 1.760  & --    &   --    &  --     \\
\hline
$e^-$-Kr & 0   & 20.79 & 1.760  & 56.84 & 6  & 0   \\
         & 1   & 20.79 & 1.760  & 97.80 & 6  & 0   \\
         & 2   & 18.25 & 1.317  &    0  & -- & --      \\
         & 3   & 14.80 & 1.409  &    0  & -- & --      \\
         & 4   & 16.98 & 1.544  &    0  & -- & --
\end{tabular}
\end{ruledtabular}
\end{table}

\begin{figure}[ht!]
\includegraphics[width=10cm]{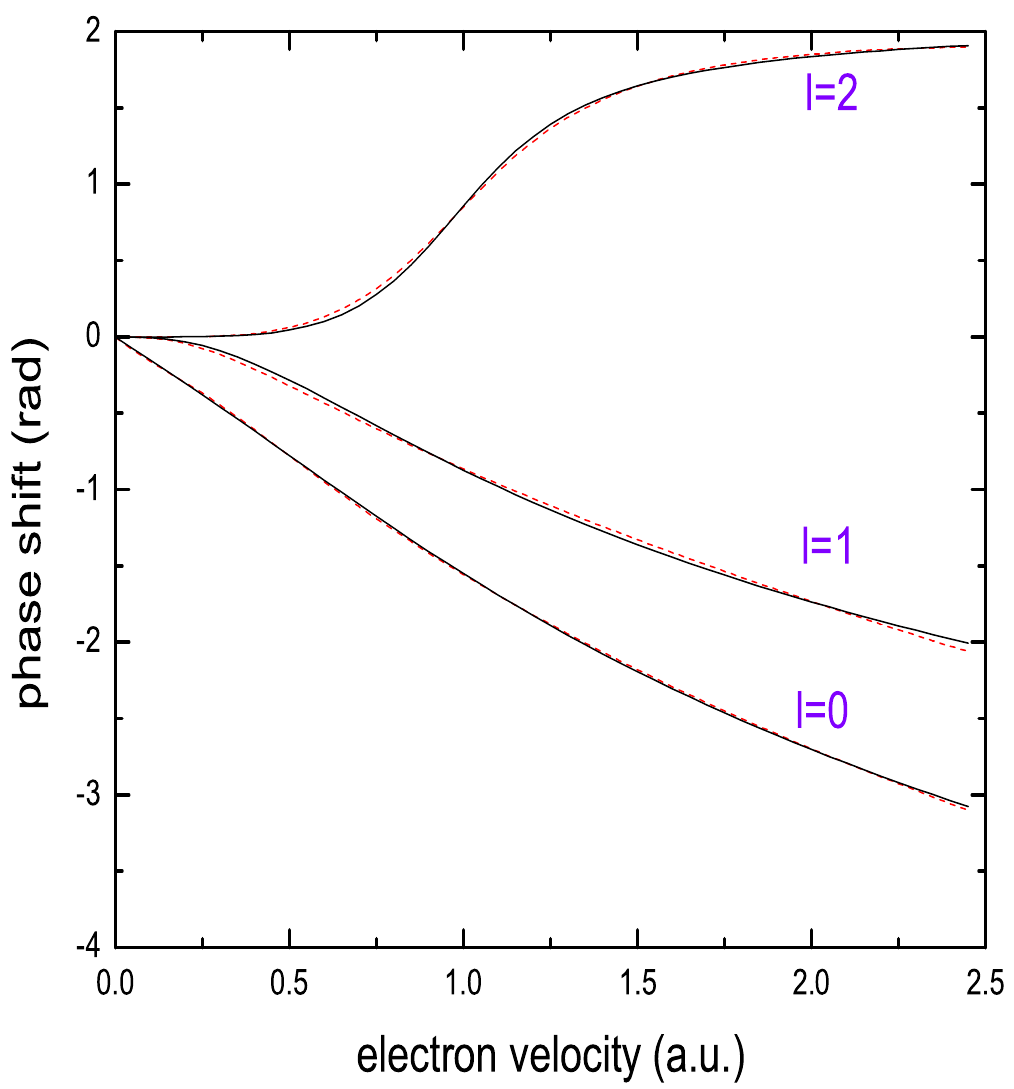}
%\picturelandscape{0}{el_Ar.eps}
\caption{Electron scattering phase shifts (modulo $\pi$)
for Ar obtained using the static-exchange (i.e., Hartee-Fock) atomic
potential (solid black lines), and the pseudopotential (\ref{V_p}) 
(dashed red lines), with parameters listed
in Table~\ref{tb:parameters}.}
\label{fig:elAr}
\end{figure}

For the electron, the pseudopotential is optimized separately for
each partial wave, by fitting the Hartree-Fock scattering phase shifts.
The corresponding sets of parameters for Ar and Kr are given in
Table~\ref{tb:parameters}, and the phase shifts for Ar are
shown in Fig.~\ref{fig:elAr} for $l=0$--2.
The pseudopotential for the $s$ wave gives the phase shifts that are almost 
indistinguishable from the Hartree-Fock, and the fitted phase shifts for the 
$p$ and $d$ waves are also quite accurate.
The behavior of the positron and electron phase shifts and the quality of the 
pseudopotential fits for Kr are similar to those shown
for Ar. For both atoms, the $d$-wave displays a broad
resonance at the electron velocity $v\sim 1$~a.u. (see Fig.~\ref{fig:elAr}).

\section{Results and discussion}

In this section, we present the Ps-atom scattering phase shifts calculated
from Eq.~(\ref{eq:Sch}) which contains the static Ps pseudopotential, and
with the inclusion of the van der Waals interaction (\ref{vdW}). We also use
these phase shifts to compute the Ps-atom elastic scattering cross section,
and compare our results with experiment and other theories.

\subsection{Ps-Kr scattering phase shifts}\label{subsec:PsKr_phase}

\begin{figure}
\centering
\includegraphics[width=10cm]{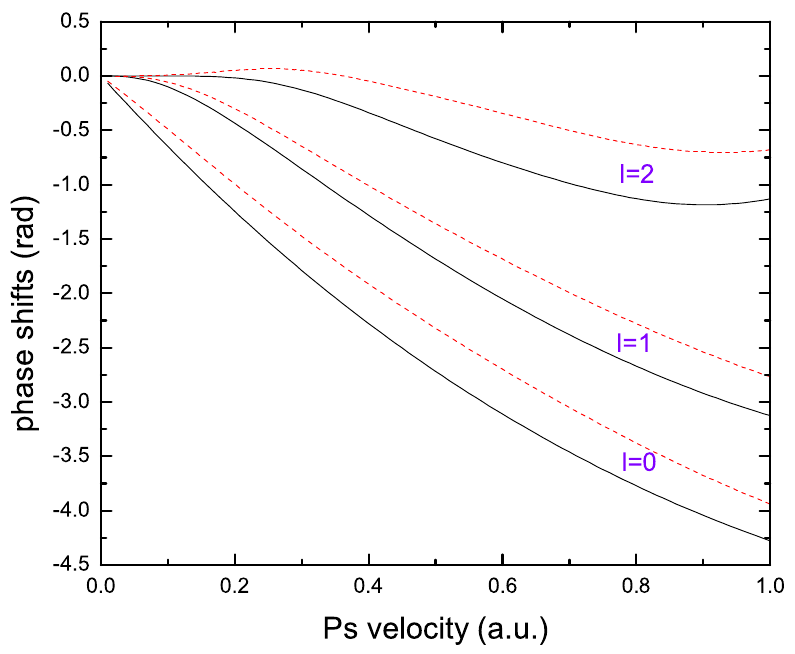}
%\picturelandscape{0}{phases_kr.eps}
\caption{(Color online) Ps-Kr scattering phase shifts for $L=0$, 1 and 2.
For each partial wave, the lower curve (solid black) is for the static
pseudopotential, and the upper curve (dashed red) includes the effect of
the van der Waals interaction with $C_6=152$~a.u. and $R_c=3.0$~a.u.}
\label{fig:krypton_1}
\end{figure}

Figure \ref{fig:krypton_1} shows the phase shifts for Ps with the orbital
angular momentum $L=0$, 1, and 2, scattered from Kr.
Adding the van der Waals interaction (with $R_c=3.0$
a.u., see Sec.~\ref{subsec:PsKr_cross}) produces relatively small positive
corrections for the $s$ and $p$ waves, while for the $d$ wave the
correction is relatively large.
In particular, due to the van der Waals potential, the phase shift acquires 
a characteristic rise, $\delta _2\propto k^4$, at low Ps momenta $k$ [see
Eq.~(\ref{eq:delvdW})].

\begin{figure}
\centering
\includegraphics[width=10cm]{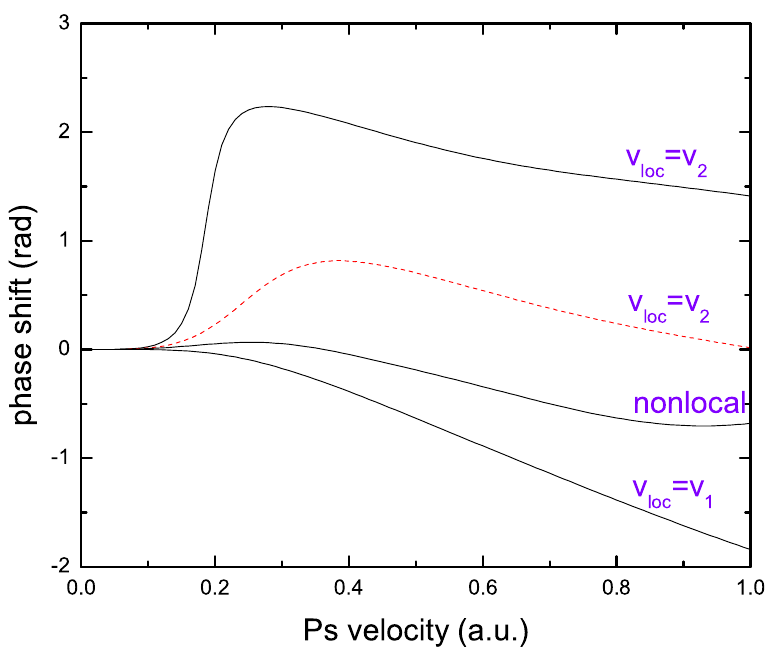}
%\picturelandscape{0}{phases_kr2.eps}
\caption{(Color online) Ps-Kr scattering phase shifts for $L=2$.
% GG 14/09 Some adjustments in the caption.
Shown are the results obtained with the full pseudopotential (i.e., with 
nonlocal effects) and its local approximations $v_{\rm loc}(r)=v_{l}(r)$ for
$l=1$ and 2. The van der Waals interaction is included in all 
cases. Solid black lines: $R_c=3.0$~a.u.,
red dashed line: $R_c=3.5$~a.u.
}
\label{fig:krypton_2}
\end{figure}

% GG 14/09 Some change of wording below.
Our calculations show that for $L=0$ and 1, the nonlocal effects related
to the second term on the right-hand side of Eq.~(\ref{Vtilde_Ps}) are small, 
but for $L=2$ they are
significant. This can be seen from Fig.~\ref{fig:krypton_2}, which compares
the full (nonlocal) $d$-wave phase shift with those obtained using two 
choices of the local potential, namely, $v_{\rm loc}(r) =v_1(r)$ and
$v_{\rm loc}(r) =v_2(r)$, where $v_{l}(r)$ is defined by Eq.~(\ref{vl_1}).
The choice of $v_{\rm loc}=v_2$ leads to a shape resonance due to the 
combination of the van der Waals interaction and the centrifugal barrier. 
The resonance becomes very pronounced at smaller $R_c$.
However, for $v_{\rm loc}=v_1$, the resonance is not visible,
and the result obtained with the full potential (i.e., including the local 
and nonlocal terms) confirms that
the nonlocal effects suppress the resonance behavior of the Ps $d$ wave.
This suppression leads to a better agreement with the static-exchange
calculations of Blackwood {\it et al.} \cite{Bla02} (see below).
A trace of this resonance is a weak minimum in the $L=2$ partial
cross section (see Sec.~\ref{subsec:PsKr_cross} below).

Analysis of the Ps $s$-wave phase shift at low momenta ($\delta _0\simeq -Ak$) 
yields the value of the scattering length $A$. In the static approximation we 
find $A=3.32$ a.u., and when the van der Waals interaction is included
we obtain $A=2.35$~a.u. (for $R_c=3.0$), or $A=2.50$~a.u. (for $R_c=3.5$).
These values can be compared
with those of Mitroy and Bromley \cite{Mit03}, i.e., $A=3.18$~a.u. in the static
approximation, and $A=1.98$~a.u. with the van der Waals interaction included. 
Note that because of the uncertainty in the van der Waals
interaction effect in Ref.~\cite{Mit03}, the corresponding scattering length 
varies between 1.22 and 2.26~a.u., with $A=1.98$~a.u. being their best 
prediction.

\subsection{Effect of the van der Waals interaction on the scattering length}

As seen from the $s$-wave phase shift in Fig.~\ref{fig:krypton_1}, the effect
of the van der Waals interaction is not very significant. It does not change
the sign of the scattering length (the way atomic polarization does for the
electron and positron scattering), and there is no Ramsauer-Townsend minimum 
in the cross section. To understand this qualitatively and semiquantitatively, 
consider a model potential with a hard repulsive core
of radius $R_0$ and a van der Waals ``tail'',
\begin{equation}\label{hw+vdW}
V(R)=
\begin{cases}
+\infty  ,& R<R_0 \\
-C_6/R^6  ,& R>R_0
\end{cases}.
\end{equation}
Here the repulsive core mimics the exchange interaction between the electron 
and the atom. The $s$-wave radial Schr\"{o}dinger equation for $k=0$ at $R>R_0$
is
\[
\frac{d^2f_0}{dR^2}+\frac{2 m C_6}{R^6}f_0(R)=0.
\]
Its solution (up to a normalization constant) is
\begin{equation}\label{eq:f0}
f_0(R)=R^{1/2}\left[J_{-1/4}(x_0)J_{1/4}(x)-J_{1/4}(x_0)J_{-1/4}(x)\right],
\end{equation}
where
\begin{equation}\label{eq:x0}
x=\frac{\sqrt{m C_6/2}}{R^2},\quad  x_0=\frac{\sqrt{m C_6/2}}{R_0^2},
\end{equation}
and $J_{\nu}$ is the Bessel function. The scattering length, obtained from
the asymptotic behavior of Eq.~(\ref{eq:f0}) at $R\to\infty$,
$f_0(R)\propto R-A$, is
\begin{equation}\label{eq:A}
A=\left(\frac{ m C_6}{8}\right)^{1/4}\frac{\Gamma(3/4)}{\Gamma(5/4)}
\frac{J_{-1/4}(x_0)}{J_{1/4}(x_0)}=R_0\left( 1-\frac{2x_0^2}{15}-\frac{22\,x_0^4}{1575}-\frac{844\,x_0^6}{482625}+\dots \right).
\end{equation}
Here the factor $(mC_6/8)^{1/4}\Gamma(3/4)/\Gamma(5/4)$ is similar to the mean 
atom-atom scattering length determined by the long-range part of the van der 
Waals interaction (see Ref.~\cite{GF93}).

\begin{figure}[ht!]
\centering
\includegraphics[width=10cm]{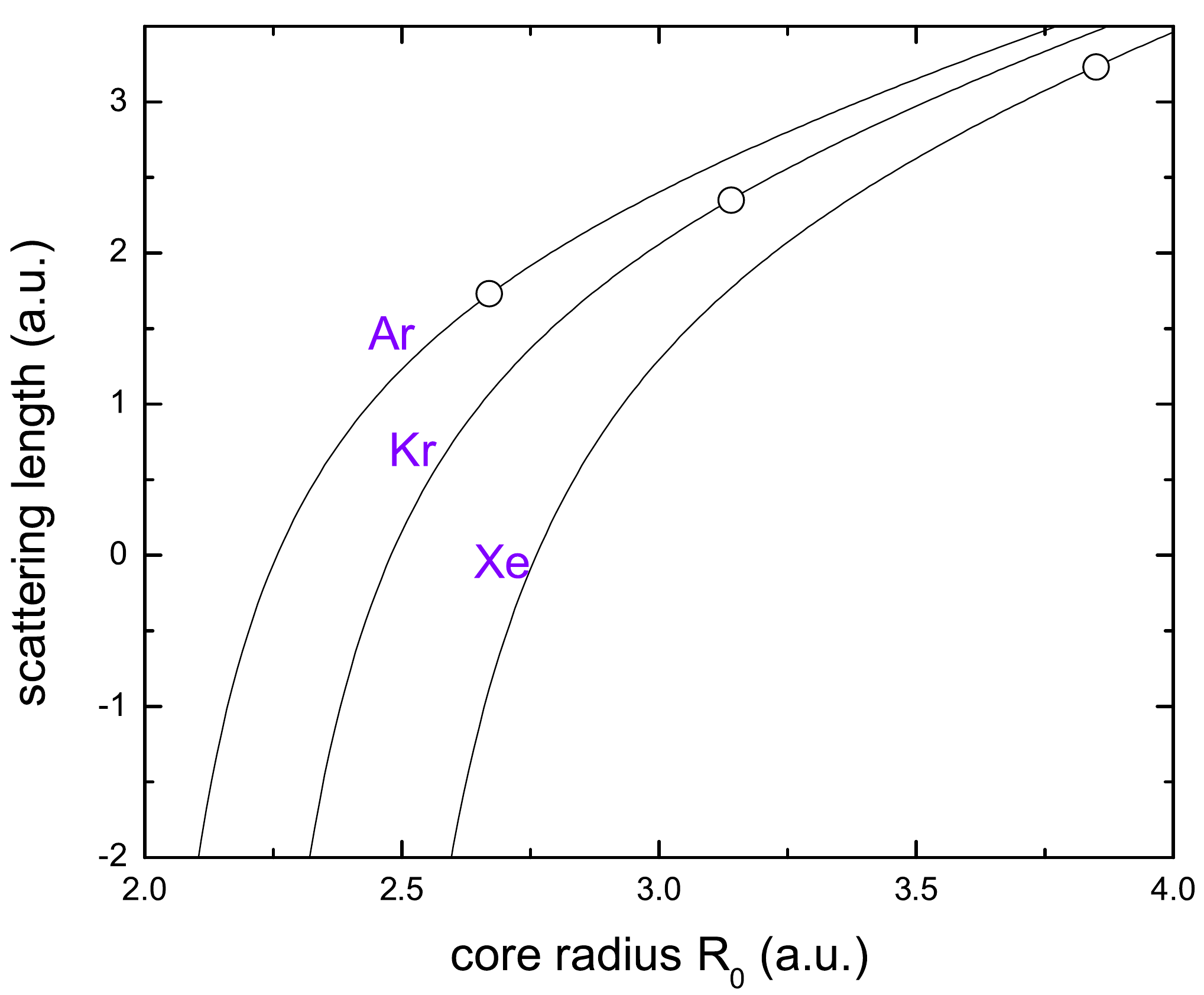}
%\picturelandscape{0}{sc_length.eps}
\caption{Dependence of the scattering length, given by Eq.~(\ref{eq:A}), for
the van 
der Waals potential (\ref{hw+vdW}) on the radius of the repulsive core. The
three curves correspond to $C_6=104.5$, 152, and 234 a.u., for Ps-Ar, Ps-Kr, 
and Ps-Xe, respectively. Open circles show the estimates of $A$ from 
Table \ref{tb:sc_length}.}
\label{fig:sc_length}
\end{figure}

Figure \ref{fig:sc_length} shows the scattering length (\ref{eq:A}) as a
function of $R_0$ for three values of the van der Waals constant, $C_6=104.5$, 
152, and 234 a.u., corresponding to Ps-Ar, Ps-Kr, and Ps-Xe interactions. 
The scattering length becomes
negative only for unrealistically small values of $R_0$.
As an estimate, we can assume that $R_0$ equals the scattering length
in the static approximation. For Kr, this gives $R_0=3.32$ a.u.
Using this value in Eq.~(\ref{eq:A}), we see that the van der Waals force 
reduces the scattering length to $A=2.67$~a.u.,
in good accord with the calculations (Sec.~\ref{subsec:PsKr_phase}).
Therefore, the effect of the van der Waals interaction is not as drastic
as the effect of the polarization interaction in $e^-$-Kr (or $e^+$-Kr)
scattering, where it makes the scattering length negative. Obviously this 
is due to the shorter range of the van der Waals force as compared to the
polarization force.

Alternatively, one can estimate $R_0$ using the mean radius of the outer atomic
orbital $\langle r\rangle $, e.g., as
\begin{equation}\label{eq:R0}
R_0=\gamma\langle r\rangle ,
\end{equation}
where $\gamma \sim 1$ is a dimensionless factor. For Kr,
$\langle r\rangle =1.95$~a.u. \cite{Rad85}, and in order to obtain the correct 
scattering length, $A=2.35$ a.u. (for $R_c=3.0$),
we should choose $\gamma=1.61$, which leads to $R_0=3.14$ a.u., close to our 
previous estimate of $R_0=3.32$ a.u.

We can use this simple model to estimate the Ps scattering lengths for other 
atoms. Table~\ref{tb:sc_length}
lists the values of $\langle r\rangle$ from Ref.~\cite{Rad85} and the 
corresponding values of the model scattering
length (\ref{eq:A}) for Ar, Kr and Xe, obtained using $\gamma =1.61$. Also 
shown are the results of the scattering
calculations of this work and of Mitroy {\it et al.} \cite{Mit01,Mit03}.
%CORRECTED table, 9/4/14
\begin{table}[ht!]
\begin{ruledtabular}
\caption{Mean atomic radii $\langle r\rangle$, core radii $R_0$, and Ps-atom 
scattering lengths $A$ for Ar, Kr and Xe. All values are in a.u.}
\label{tb:sc_length}
\begin{tabular}{ccccccc}
System & $C_6$ & $\langle r\rangle$ & $R_0$ & $A$\footnote{Scattering length 
from Eq.~(\ref{eq:A}), obtained using Eq.~(\ref{eq:R0}) with $\gamma =1.61$.} 
& $A$\footnote{Present scattering calculations.} & $A$\footnote{Values 
obtained by Mitroy \textit{et al.} \cite{Mit01,Mit03}.}\\
\hline
Ar & 104.5 & 1.66 & 2.67 & 1.73  & 2.14--2.33 & 1.30--1.98  \\
Kr & 152   & 1.95 & 3.14 & 2.35  & 2.35--2.50 & 1.22--2.26  \\
Xe & 234   & 2.39 & 3.85 & 3.23  &    --      & 1.50--2.60
\end{tabular}
\end{ruledtabular}
\end{table}

Although the model result for Xe looks somewhat overestimated, the model
correctly predicts the trend of the scattering length $A$ to increase with 
the atomic number $Z$. Although the van der Waals interaction (which makes 
$A$ smaller) grows with $Z$, the increase of the atomic radius, i.e., the 
effect of the static repulsion, is stronger. [This can be seen from the 
expression for $x_0$, given by Eq.~(\ref{eq:x0}), which is proportional to
$C_6^{1/2}$ but inversely proportional to $R_0^2$.]
This is opposite to what is observed in electron-atom scattering
where the effect of the polarization attraction takes over the effect of the 
core radius, and the scattering length decreases with the growth of
$Z$. (The analog of the parameter $x_0$ in this case is proportional to 
$\sqrt{\alpha }/R_0$, see Ref.~\cite{GF93}.)

These observations have  an important consequence for the
comparison of $e^-$-$A$ scattering with Ps-$A$ scattering. The
observed similarities \cite{Bra10} at energies above the Ps
ionization threshold are explained in terms of the impulse
approximation \cite{Fab14}. On the other hand, in the region below
the ionization threshold, where the impulse approximation fails,
no similarity exists. In this energy range scattering is
controlled by different long-range interactions, the strong
polarization interaction for the electrons, and the relatively
weak van der Waals interaction in the case of Ps.

\subsection{Ps-Kr scattering cross section}\label{subsec:PsKr_cross}

\begin{figure}[ht!]
\centering
\includegraphics[width=10cm]{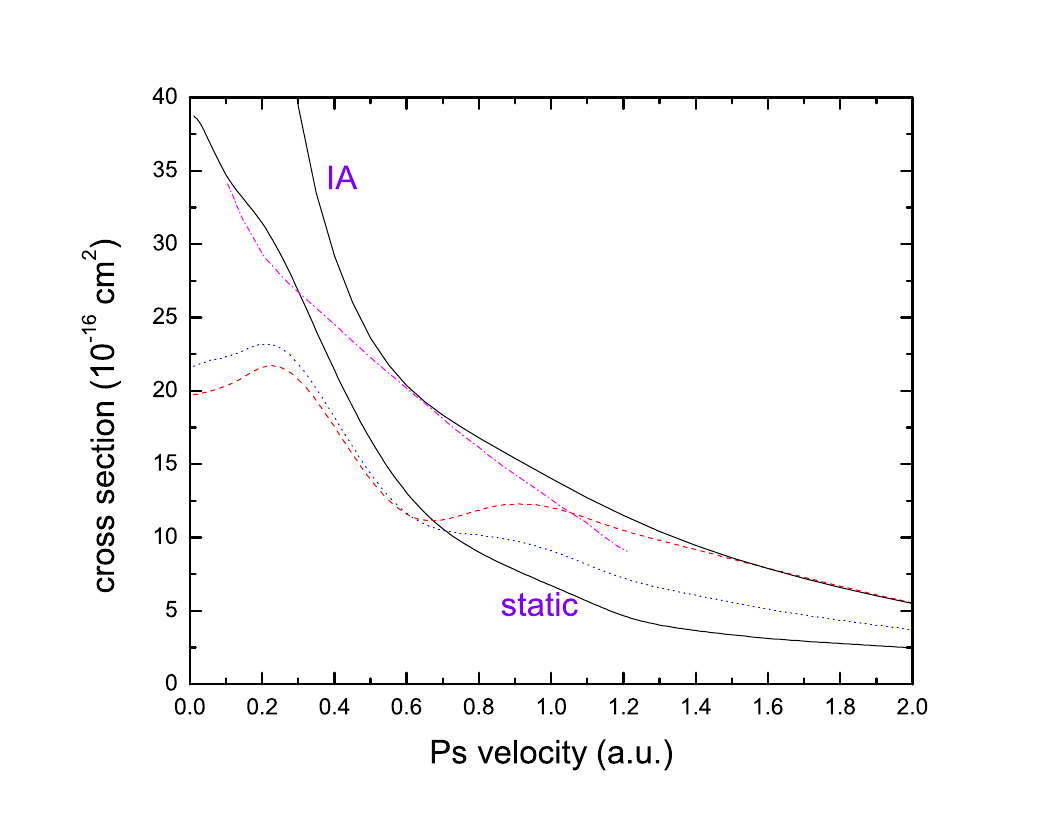}
%\picturelandscape{0}{elastic_Kr.eps}
\caption{(Color online) Ps-Kr elastic scattering cross sections. Solid black
curve ``IA'' is the result of the impulse approximation \cite{Fab14};
solid black curve ``static'' is the present static-field calculation
(i.e., $C_6=0$); dashed red curve is the present calculation with the static
and van der Waals interaction ($C_6=152$ a.u., $R_c=3.0$ a.u.); dotted blue
curve is the same for $R_c=3.5$ a.u.; dot-dashed magenta curve is the
static-exchange calculations of Blackwood {\it et al.} \cite{Bla02}.
}
\label{fig:krypton_crs1}
\end{figure}

Figure~\ref{fig:krypton_crs1} shows the elastic Ps-Kr cross sections in
the velocity range from threshold to 2~a.u. Higher partial waves ($L>4$),
up to $L=10$,
were included by solving the radial equation in the local approximation
with inclusion of the van der Waals interaction.
The van der Waals interaction was included with two choices of
the cut-off parameter: $R_c=3.0$ and 3.5~a.u. The figure also shows
the cross section obtained by Blackwood {\it et al.} \cite{Bla02} using the
static-exchange approximation, i.e., without inclusion of virtual excitations
in the target or projectile. These results can be compared with our static
calculations. Figure \ref{fig:krypton_crs1} shows that the two theories are
very close at low velocities, but the present static-field cross section
decreases faster with the increasing velocity, compared to that of
Blackwood {\it et al.} However, after inclusion of the van der Waals
interaction, our cross section increases significantly at $v>0.6$~a.u.,
and merges with the result of the impulse approximation for $R_c=3.0$ a.u.
We therefore choose this value of $R_c$ for comparison with the experiment
(see below).

At low velocities (below the Ps ionization threshold, $v=0.5$~a.u.), the van
der Waals interaction leads to a significant reduction of the elastic cross
section.
This effect is due to the decrease in the scattering length and a general
increase of the low-$l$ phase shifts (i.e., decrease in absolute magnitude),
as seen in Fig.~\ref{fig:krypton_1}.
The van der Waals interaction also results in two
features in the cross section: a local maximum at $v=0.22$~a.u. 
and a local minimum at $v=0.67$~a.u. (for $R_c=3.0$~a.u.). The former is due 
to the $s$-wave contribution and
is caused by the long-range attractive interaction. To understand this, it is 
useful to discuss a similar effect in the low-energy electron-atom scattering 
which is controlled by the polarization interaction.
According to the modified effective-range theory of O'Malley {\it et al.}
\cite{OMa61}, at low energies the $s$-wave phase shift behaves as
\begin{equation}\label{eq:OM}
\tan\delta_0=-Ak-\pi k^2\alpha /3+O(k^3\ln k),
\end{equation}
where $\alpha $ is the atomic polarizability. Here the
characteristic \textit{quadratic} part of the polarization contribution to the
phase shift [second term in Eq.~(\ref{eq:OM})] is negative (although the total
contribution of the attractive polarization potential is positive). As a
result, for a negative scattering length $A$, the phase shift passes through
zero at small $k>0$, leading to the Ramsauer-Townsend effect. In contrast,
for $A>0$ the phase shift decreases faster than linear, which gives rise to 
a maximum in the partial cross section. For example, a
maximum is observed in electron scattering from Ne \cite{McE83}, for which
the scattering length is small and positive. (Note that although the maximum
in the {\it total} cross section for Ne is observed at about $E=25$ eV, the
$s$-wave contribution peaks at $E=6.7$ eV, still quite a large energy compared
to the position of a typical Ramsauer-Townsend minimum.)
One could call this phenomenon the ``anti-Ramsauer'' effect, although we are
not aware of the use of such term in the literature.

A similar situation occurs in Ps-atom scattering, although now the
additional contribution to the phase shift comes from the van der Waals
interaction. As shown in Appendix~\ref{app:effrange}, the modified
effective-range expansion of the $s$-wave scattering phase shift reads as
\begin{equation}
\tan\delta_0=-Ak-Bk^3+2 m C_6\pi k^4/15 +O(k^5),
\label{MERT_1}
\end{equation}
where $B=\frac{1}{2}r_0A^2$, and $r_0$ is the effective range \cite{LL}. The 
coefficient $B$ depends on both the short-range and the long-range (van der 
Waals) interactions. If $B>0$, the major correction to the $-Ak$ behavior is 
negative,
since the $k^4$ term is relatively small at low energies.
Our calculations show that this is indeed the case, although the expansion 
(\ref{MERT_1}) is valid only at very low energies. Hence, a weak
anti-Ramsauer effect is observed. Naturally, the effect is
small compared to that observed in electron-atom scattering because of the
relative weakness of the van der Waals interaction compared to the polarization interaction. 

The minimum at $v=0.67$ a.u. observed in Fig.~\ref{fig:krypton_crs1} for
$R_c=3.0$~a.u. is due to the $d$-wave shape resonance, which is quite
pronounced in the $e^-$-Kr scattering \cite{McE84}.
However, in Ps-Kr scattering, this resonance is suppressed, as discussed in Sec.~\ref{subsec:PsKr_phase} and seen in Figs. \ref{fig:krypton_1} and \ref{fig:krypton_2}. Due to a strong
background contribution, this resonance appears as a window.

\begin{figure}[ht!]
\centering
\includegraphics[width=10cm]{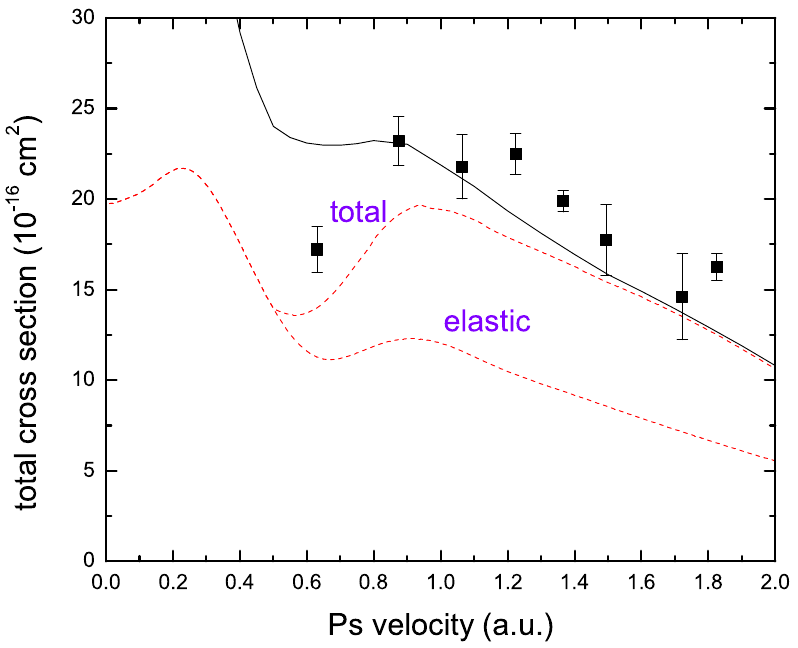}
%\picturelandscape{0}{total_kr.eps}
\caption{(Color online) Ps-Kr total scattering cross sections. Solid black
curve is the total cross section in the
impulse approximation \cite{Fab14};
dashed red curves present elastic and total cross sections obtained
with $C_6=152$ a.u., $R_c=3.0$ a.u. Solid squares are experimental data from
Ref.~\cite{Bra10}.}
\label{fig:krypton_crs2}
\end{figure}

To compare with the experimental Ps-atom total scattering cross section,
the Ps ionization cross section should be added to the elastic cross section.
Indeed, Ps ionization contributes significantly at velocities $v>0.5$~a.u.
As in the
impulse-approximation calculations \cite{Fab14}, the ionization cross
sections are taken from Ref.~\cite{Sta05}. Figure~\ref{fig:krypton_crs2} 
shows the elastic and total cross section computed in the present
work together with the results of the impulse approximation and experiment. 
Although the experimental data \cite{Bra10} are not available at low 
velocities, the data point at $v=0.63$ a.u. indicates that the cross section 
should slightly decrease towards lower velocities. This trend is confirmed by 
our results.
The peaking of the experimental cross section at
$v\approx 0.9$ a.u. is also in agreement with our results. In addition, our 
calculation
predicts a weak local maximum at $v=0.23$~a.u. and a local minimum at
$v=0.56$~a.u. Both of these predictions call for experimental verification.

\subsection{Ps-Ar scattering}

\begin{figure}[ht!]
\centering
\includegraphics[width=10cm]{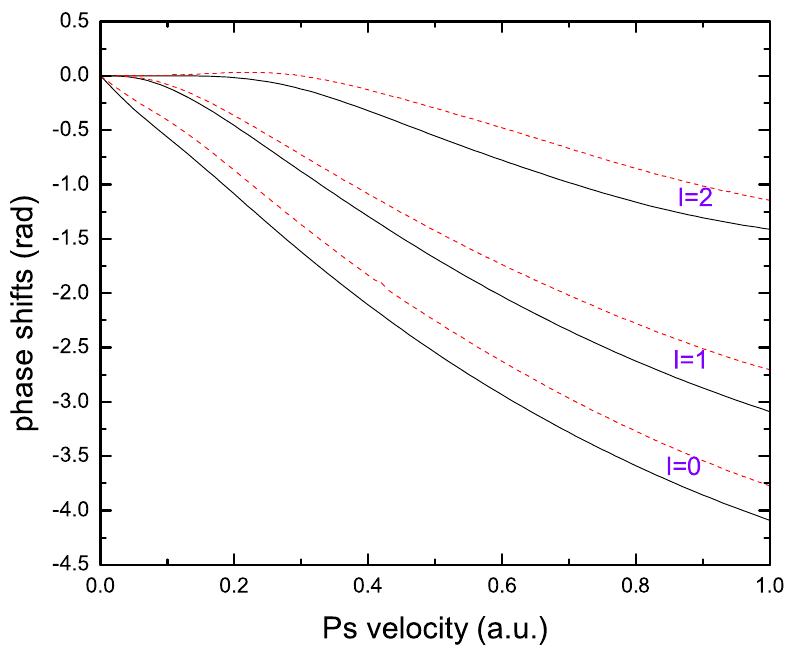}
%\picturelandscape{0}{phases_ar.eps}
\caption{(Color online) Ps-Ar scattering phase shifts. For each $L$, the lower
curve (solid black) was calculated using the static-field pseudopotential,
while the upper (dashed red) also includes the effect of
the van der Waals interaction with $C_6=104.52$~a.u. and $R_c=3.0$~a.u.
}
\label{fig:argon_1}
\end{figure}

\begin{figure}[ht!]
\centering
\includegraphics[width=10cm]{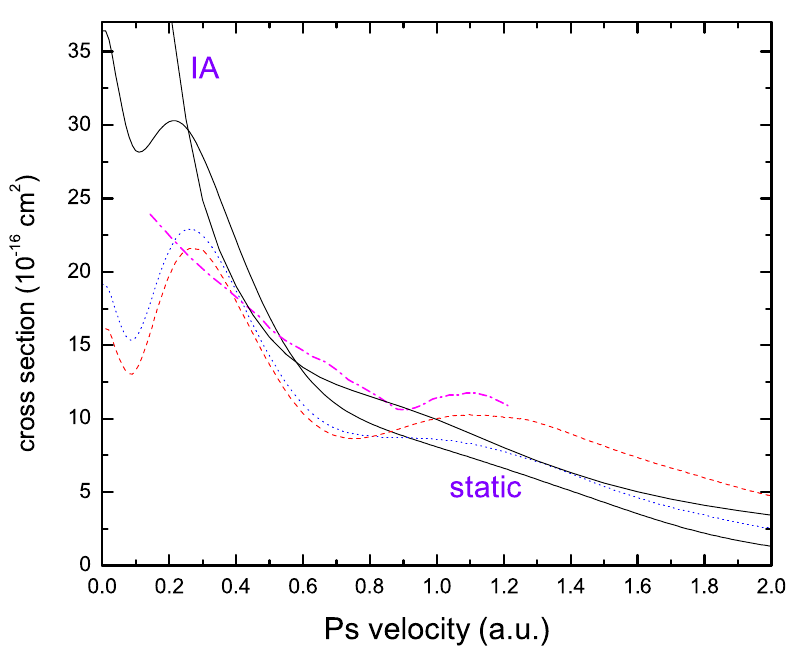}
%\picturelandscape{0}{elastic_ar.eps}
\caption{(Color online) Ps-Ar elastic scattering cross sections. Solid black 
curve ``IA'' is the result of the impulse approximation \cite{Fab14}; solid
black curve ``static'' is the present static calculation ($C_6=0$); dashed 
red curve is the calculation with the static and van der Waals potential
($C_6=104.5$ a.u., $R_c=2.5$ a.u.); dotted blue curve is the same for
$R_c=3.0$ a.u.; dot-dashed magenta curve is the close-coupling
calculations of Blackwood {\it et al.} \cite{Bla02}.}
\label{fig:argon_crs1}
\end{figure}

Figure \ref{fig:argon_1} shows the $s$-, $p$- and $d$-wave phase shifts for
Ps-Ar scattering. All of the main features here are the same as in the Ps-Kr
scattering. However, $\tan\delta_0$ follows
the linear $-Ak$ behavior much more closely. 
% GG 15/09 Replaced "contribution" by "cross section", mention small r_0.
This results in a decrease of the $s$-wave cross section, according to
\[ \sigma_0\simeq 4\pi\frac{A^2}{1+A(A-r_0)k^2}, \]
where $r_0$ is, in fact, quite small.
At the same time the $p$-wave contribution increases rather sharply from
threshold, leading to a local maximum in the total cross section at
$v=0.22$~a.u. in the static approximation ($C_6=0$),
and an even more pronounced local maximum at $v=0.26$~a.u. when
the van der Waals interaction is included. The total elastic cross sections 
are shown in Fig.~\ref{fig:argon_crs1}.

The local maximum in the total cross section is controlled by the local part 
of the Ps-Ar pseudopotential for $L=0$ and $L=1$. Since this potential is 
very sensitive to the partial cancellation of the attractive part due to the 
$e^-$-Ar interaction
and the repulsive part due to the $e^+$-Ar interaction, the position and the
magnitude of the maximum is subject to uncertainties. It is possible that
effects not included in the present calculations, e.g.,
short-range correlations, can change the position and shape of the
maximum or even eliminate it completely. We note that the close-coupling
calculations of Blackwood {\it et al.} \cite{Bla02}, which allow for the
virtual excitations of the Ps, but keep the target atom frozen, do not exhibit 
a low-energy maximum at all. However, they do show a local maximum at 
$v=1.1$~a.u., which in our calculation is due to the $d$-wave contribution.
Overall, Fig.~\ref{fig:argon_crs1} shows that there is a
reasonable agreement with calculations \cite{Bla02} for $v=0.4$--1.2~a.u.. 
One can also see that the effect of the
van der Waals potential for Ar is smaller than it is for Kr
(Fig.~\ref{fig:krypton_crs1}), and the cross
section is not as sensitive to the choice of $R_c$ (which can be related to
the smaller $C_6$ value for Ar). For a more detailed comparison between the
present calculation and that of
Blackwood {\it et al.} \cite{Bla02}, information
on partial cross sections would be needed.

The Ps-Ar scattering length in the static approximation is $A=3.19$~a.u., 
whereas the value calculated by Mitroy and Ivanov \cite{Mit01} is 2.85~a.u. 
After adding the van der Waals interaction we obtain $A=2.33$~a.u. for 
$R_c=3.0$~a.u.,
and $A=2.14$~a.u. for $R_c=2.5$~a.u. The corresponding value obtained by
Mitroy and Ivanov varies between 1.30 and 1.98~a.u. Comparing with the Ps
scattering lengths obtained for Kr (Sec.~\ref{subsec:PsKr_phase}), we
see that the values for both atoms are quite close. While the Ar-Kr system
has a smaller $C_6$ value, the Ar atom has a smaller radius, and the two
effects largely compensate for each other (see Fig.~\ref{fig:sc_length} and 
Table \ref{tb:sc_length}).

\begin{figure}
\centering
\includegraphics[width=10cm]{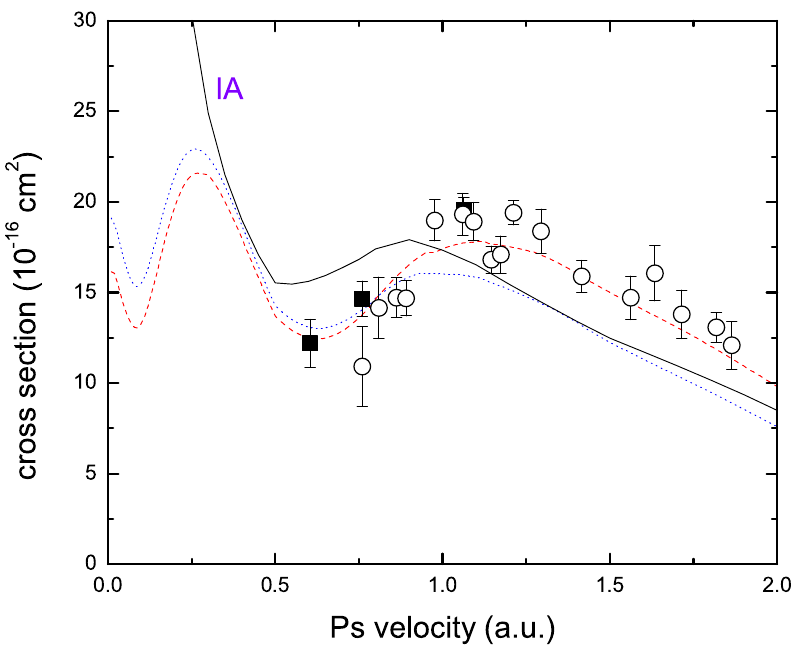}
%\picturelandscape{0}{total_ar.eps}
\caption{(Color online) Ps-Ar total scattering cross sections. Solid black
curve ``IA'' is the result of the impulse approximation \cite{Fab14};
dashed red curve is the present elastic cross section obtained with
$C_6=104.5$~a.u., $R_c=2.5$~a.u., augmented by the ionization cross sections of
Starrett {\it et al.} \cite{Sta05}; dotted blue curve is the same for
$R_c=3.0$~a.u. Open circles: experiment \cite{Gar00}; solid squares:
experiment \cite{Bra10}.}
\label{fig:argon_crs2}
\end{figure}

As seen in Fig. \ref{fig:argon_crs1}, the choice of $R_c=3.0$~a.u. matches
better with the impulse approximation results at higher velocity. However, the
smaller value of $R_c=2.5$~a.u. leads to a better agreement with the
experiment. Figure~\ref{fig:argon_crs2} shows the total
cross sections for Ps-Ar scattering obtained for each of these cut-off radii.
Whereas both theoretical curves describe well the overall behavior of the
measured cross sections \cite{Gar00} at $v>0.5$~a.u., the cross section 
obtained with
$R_c=2.5$ a.u. agrees better with the measured absolute values. This value
also appears to be more physical, as it is smaller than the optimal cut-off
radius of $R_c=3.0$~a.u. found for Kr.

\section{Conclusion}
The pseudopotential model developed in the present paper describes
Ps scattering from noble-gas atoms (Ar and Kr) at energies below
the ionization threshold and matches the impulse-approximation results above
the ionization threshold. Although experimental data are not available in the
low-energy region, our results describe well the trend seen in the
experimental cross sections to drop with decreasing velocity below
$v\approx 1$~a.u. In addition, our calculations predict zero-energy
cross sections (or scattering lengths) which are in accord with stochastic
variational calculations \cite{Mit01,Mit03}.

Analysis of the scattering phase shifts shows that the static Ps-atom
interaction is repulsive. This repulsion arises from the electron Pauli
exclusion from closed-shell atoms (while the pure electrostatic interaction
is zero for the truly neutral Ps atom).
The phase shifts also indicate that the role of correlations represented by
the van der Waals interaction at low energies, is relatively small.

Because of the relative weakness of the van der Waals interaction compared to
the polarization interaction in electron-atom scattering, the scattering
lengths for both Ar and Kr are positive, and the Ramsauer-Townsend minimum
is not observed for these targets. The overall picture of Ps-$A$ scattering is 
quite different from the $e^-$-$A$ scattering in the low-energy region. This 
is in stark contrast to the intermediate energy range from the Ps ionization
threshold up to $v\sim 2$~a.u. Here the Ps-$A$ scattering is mostly controlled
by the electron-atom exchange, which makes its cross section very similar to
that for $e^-$-$A$ scattering. In the low-energy region, where the long-range
interaction is important (especially for the electrons), this similarity
disappears.

Although the van der Waals interaction in Ps-$A$ scattering does not produce
the Ramsauer-Townsend minimum, it can lead to more subtle features in the
cross sections, such as low-energy maxima. However these
features are subject to uncertainties because of a delicate balance between
the repulsive and attractive components of the Ps-$A$ interaction. They can
also be affected by other effects such as short-range correlations, which are
not included explicitly in the present calculation. The uncertainty can be
resolved by performing accurate measurements of Ps-$A$ scattering at low
energies, and by new fully correlated calculations for this
interesting and challenging system.

\section*{Acknowledgments}
The authors are grateful to G. Laricchia for stimulating discussions.

\appendix

\section{Expansion of the Ps ground state density}\label{app:F}

We start with a known expression for the free-particle Green's function
(see, e.g., \cite{MVK}),
\[
\frac{e^{ikr_{12}}}{r_{12}}=ik\sum_{l=0}^\infty(2l+1)h_l^{(1)}(x_>)j_l(x_<)P_l
(\cos\theta_{12}),
\]
where $x_>=kr_>$, $x_<=kr_<$, and $j_l$, $h_l^{(1)}$ are the spherical Bessel
and Hankel functions.

To switch to the decaying exponent, we make a substitution $k=i\kappa$
and arrive at the following expansion
\begin{equation}\label{exp_exp}
e^{-\kappa r_{12}}=\sum_{l=0}^\infty (2l+1)F_l(r_1,r_2)P_l(\cos\theta_{12})
\end{equation}
where
\[
F_l(r_1,r_2)=\frac{d}{d\kappa}\left[\kappa h_l^{(1)}(i\kappa r_>)
j_l(i\kappa r_<)\right].
\]
It is convenient now to introduce the following real functions:
\[
\hat{h_l}(x)=-i^lh_l^{(1)}(ix),\quad \hat{j_l}(x)=i^lj_l(ix),
\]
which are related to the modified Bessel function $K_{l+1/2}(x)$ and
$I_{l+1/2}(x)$. Explicit expressions for the first few of these functions are
\[
\hat{h}_0(x)=\frac{e^{-x}}{x},\quad \hat{h}_1(x)=\frac{e^{-x}}{x}
\left(1+\frac{1}{x}\right),\quad \hat{h}_2(x)=\frac{e^{-x}}{x}
\left(1+\frac{3}{x}+\frac{3}{x^2}\right)
\]
\[
\hat{j}_0(x)=\frac{\sinh x}{x},\quad
\hat{j}_1(x)=\frac{1}{x}\left(\frac{\sinh x}{x}-\cosh x\right),
\quad \hat{j}_2(x)=\frac{\sinh x}{x}\left(1+\frac{3}{x^2}\right)-
\frac{3\cosh x}{x^2}.
\]
The recurrence relations for these functions are
\[
z_{l+1}(x)=\frac{2l+1}{x}z_l(x)+z_{l-1}(x),
\]
\[
\frac{dz_l(x)}{dx}=-\frac{l+1}{x}z_l(x)-z_{l-1}(x),
\]
where $z_l$ stands for either $\hat{j}_l$ or $\hat{h}_l$. Using these, we obtain
\[
F_0(x_1,x_2)=-\hat{h}_0(x_>)\hat{j}_0(x_<)+x_>\hat{h}_1(x_>)\hat{j}_0(x_<)
+x_<\hat{h}_0(x_>)\hat{j}_1(x_<),
\]
and
\[
F_l(x_1,x_2)=(-1)^l[(2l+1)\hat{h}_l(x_>)\hat{j}_l(x_<)+
x_>\hat{h}_{l-1}(x_>)\hat{j}_l(x_<)
+x_<\hat{h}_l(x_>)\hat{j}_{l-1}(x_<)], \quad l\geq 1.
\]
In practice, expansion (\ref{exp_exp}) converges well by summing up to
$l_{\max}=30$ if, for high $l$, asymptotic expansions for $\hat{h}_l$
and $\hat{j}_l$ are used.

\section{Modified effective range expansion for the van der Waals
potential}\label{app:effrange}

We are interested in the behavior of the $s$-wave scattering phase shift
$\delta_0$. According to the effective range theory for short-range potentials, it is given by the effective-range expansion \cite{Bet49}
\begin{equation}\label{ert_1}
k\cot\delta_0=-\frac{1}{A}+\frac{1}{2}r_0k^2 +O(k^4)
\end{equation}
where $A$ is the scattering length and $r_0$ is the effective range.
Note that in the presence of a weakly bound state, $r_0>0$ \cite{LL}, but
generally this is not  true.
For small momenta and phase shifts, it is more convenient to rewrite
Eq.~(\ref{ert_1}) as
\begin{equation}\label{ert_2}
\tan\delta_0=-Ak-Bk^3+O(k^5),
\end{equation}
where $B=\frac{1}{2}A^2r_0$. More generally, $k\cot\delta_0$
in Eq. (\ref{ert_1}) can be expanded in even powers of $k$,
and $\tan\delta_0$ in Eq. (\ref{ert_2}) in odd powers of $k$.

In the presence of the long-range interaction $-C_n/r^n$ the first
``anomalous'' term in the expansion (\ref{ert_2}) is proportional to
$k^{n-2}$ \cite{LL}. This term can be calculated in the Born approximation
according to the prescription given by Landau and Lifshitz \cite{LL}. Consider
the case $n=6$ (van der Waals interaction). The corresponding correction
$\Delta f$ to the scattering amplitude is
\[
\Delta f(q)=2 m C_6q^3\int_{qR_c}^{\infty}\frac{\sin\xi}{\xi^5}d\xi ,
\]
where $q=2k\sin\theta/2$ is the momentum transfer, $\theta$ is the scattering
angle, and $R_c$ is a cut-off radius similar to that introduced in
Eq.~(\ref{vdW}). Integrating several times by parts and expanding the result
in powers of $kR_c$ at small $kR_c$ gives
\[
\Delta f(q)\approx 2 m C_6\left(\frac{1}{3R_c^3}-\frac{q^2}{6R_c}+\frac
{\pi q^3}{48}\right).
\]
Expanding this amplitude in partial waves, we obtain for the correction to
the $s$-wave phase shift:
\begin{equation}\label{ert_cor}
\Delta\delta_0=c_1k+c_2k^3+\frac{2\pi m C_6k^4}{15}
\end{equation}
where
\[ c_1=\frac{2m C_6}{3R_c^3}, \quad c_2=-\frac{2m C_6}{R_c}. \]
 The first two terms in the Eq. (\ref{ert_cor}) expression are of the same
type as those in the effective-range expansion
(\ref{ert_2}), while the last term is ``anomalous'', caused by the power-law
behavior of the potential. Therefore, the modified effective-range expansion
can be written as
\begin{equation}\label{ert_3}
\tan\delta_0=-A'k-B'k^3+\frac{2\pi m C_6k^4}{15} +O(k^5),
\end{equation}
where we have introduced the new parameters $A'$, $B'>0$ to emphasize that
they are different from those in Eq.~(\ref{ert_2}).

% GG 15/09 I replaced l by L below, as this refers to Ps orbital angular
% momentum in the context of the paper.
One might ask if this derivation is rigorous enough because of the use of
the Born approximation. In fact, the expansion in Eq.~(\ref{ert_3}) can be
derived from a more rigorous modified effective-range theory \cite{Fab79}
for the $-C_n/r^n$ potential. This theory shows that the first ``anomalous''
correction to the effective-range expansion of the phase shift can be obtained
from the analytical continuation of the integral \cite{Fab79},
\begin{equation}\label{eq:Ddelta}
\Delta\delta_L=\pi m C_nk^{n-2}\int_0^{\infty}\frac
{[J_{L+1/2}(x)]^2}{x^{n-1}}dx,
\end{equation}
which converges for $L>(n-3)/2$, to any physical value of $L$, e.g.,
$L=0$ for the $s$-wave scattering. [In (\ref{eq:Ddelta}), $J_{\nu}$ is
the Bessel function.] For $n=6$, one obtains (see also Refs.~\cite{LL,Ganas})
\begin{equation}\label{eq:delvdW}
\Delta\delta_ L=\frac{6\pi mCk^4}{(2L+5)(2L+3)
(2L+1)(2L-1)(2L-3)},
\end{equation}
which means that the lowest order correction in the effective range expansion
for $L=0$ is
\[
\Delta\delta_0=\frac{2\pi m C_6 k^4}{15},
\]
in agreement with Eq.~(\ref{ert_cor}).

For a short-range potential, the low-energy behavior of the higher
partial wave phase shifts is $\delta _L\propto k^{2L+1}$. This
means that for $2L+1>4$, i.e., $L\geq 2$, the ``anomalous''
correction (\ref{eq:delvdW}) is, in fact, the leading term in the
low-$k$ expansion. This explains the behavior of the $d$-wave
phase shifts seen in Figs.~\ref{fig:krypton_1} and
\ref{fig:argon_1} when the van der Waals interaction is included.

\end{document}